\documentclass{jfm}

\usepackage{graphicx}
\usepackage{amsmath,amssymb}
\usepackage{natbib}
\usepackage{natbib}
\usepackage{xr-hyper}
\lefttitle{F. Monroy}
\righttitle{Renormalized flow theory of wave turbulence}

\title{Renormalized flow theory of wave turbulence: Kolmogorov--Zakharov spectra as emergent asymptotic states}

\author[F. Monroy and J. A. Santiago]
{F. Monroy\aff{1,2,*}, J. A. Santiago\aff{1,3}}

\affiliation{
\aff{1}Departamento de Química Física, Universidad Complutense de Madrid, 28040 Madrid, Spain
\aff{2}Translational Biophysics, Instituto de Investigación Sanitaria Hospital 12 de Octubre (Imas12), 28041 Madrid, Spain
\aff{3}Departamento de Matemática Aplicada, Universidad Autónoma Metropolitana, Unidad Cuajimalpa, 05348 Ciudad de México, Mexico
}

\corresau{\email{monroy@ucm.es}}

\begin{document}

\maketitle

\begin{abstract}
We develop a continuous Wilsonian renormalized-flow theory of weak wave turbulence directly in spectral frequency space, for finite cascades in experimentally driven Newtonian fluids. The central quantity is a scale-dependent effective coupling that governs nonlinear transfer across logarithmic frequency shells and organizes the cascade as a finite renormalized branch. Within this formulation, the inertial interval is constructed dynamically as a plateau of the running flow, whose non-autonomous character is expressed through its explicit dependence on the logarithmic distance from the injection scale and thereby encodes the cumulative action of forcing and degradation along the cascade. The ultraviolet cutoff follows internally as the terminal scale at which the plateau branch ceases to exist, whereas the integrated spectral response is fixed by infrared matching to the injection scale. In this way, the finite inertial branch is determined by the renormalized dynamics itself, while Kolmogorov--Zakharov (KZ) spectra arise only as its asymptotic constant-flux scaling states. The theory applies to both capillary and gravity wave turbulence and admits a physically transparent realization in monochromatically driven discrete cascades, which fix the topology-dependent exponent structure of the renormalized flow.
\end{abstract}

\section{Introduction}

Wave turbulence provides a canonical description of energy transfer across scales in weakly nonlinear dispersive media \citep{NewellRumpf2011} and, for wave-bearing fluids, furnishes the standard asymptotic framework for scale-invariant spectral transport \citep{Falcon2010,FalconMordant2022}. In the broader language of turbulent cascade theory, it belongs to the statistical description of multiscale transfer in driven non-equilibrium fluid systems \citep{Frisch1995,FalkovichGawedzkiVergassola2001}. In its kinetic formulation, weak wave turbulence theory yields an evolution equation for the spectral density under the assumptions of weak nonlinearity, resonant interaction, and statistical phase mixing \citep{ZakharovFilonenko1967,ZakharovLvovFalkovich1992,NewellRumpf2011}. Within that setting, the Kolmogorov--Zakharov (KZ) spectra appear as stationary constant-flux solutions associated with scale-separated inertial regimes \citep{ZakharovLvovFalkovich1992,NewellRumpf2011}. For surface gravity--capillary waves, this kinetic framework remains the established asymptotic description of weak turbulence \citep{ZakharovFilonenko1967,DiasKharif1999,Falcon2010,FalconMordant2022}.

For finite cascades in laboratory gravity--capillary systems \citep{DiasKharif1999}, spectral transfer develops under conditions that involve finite basin size, distributed viscous dissipation, finite forcing bandwidth, and the coexistence of gravity and capillary branches, all of which shape the cascade over a bounded frequency interval \citep{Falcon2010,FalconMordant2022}. These features naturally call for a formulation directly in frequency space, where the cascade is expressed in terms of the variables most directly resolved in experiments, namely the spectral distribution of surface elevation \citep{Falcon2010,Cobelli2011,FalconMordant2022}. They also motivate a theoretical description in which the central object is the finite transfer branch itself, characterized by its formation, its extent, and its ultraviolet termination, as expected for cascades developing in finite and mesoscopic wave-turbulence settings \citep{Falcon2010,LvovNazarenko2010,FalconMordant2022}.

A particularly transparent realization of this setting is furnished by monochromatically driven cascades \citep{Falcon2010,Michel2018,FalconMordant2022}, in which energy transfer develops across a harmonic ladder organized by the forcing scale. In such systems, the forcing acts both as the source of injected activity and as the structure that selects the accessible transfer network, thereby exposing in direct form the role of interaction topology in the development of the cascade. The discrete harmonic sector thus supplies a natural ultraviolet closure for finite cascades and provides a concrete bridge between experimentally resolved spectral transport and a coarse-grained continuum description \citep{LvovNazarenko2010}. This discrete realization also furnishes the natural point of departure for the renormalized continuum formulation developed below, as progressive spectral broadening organizes transfer over increasingly connected frequency shells and prepares a scale-dependent description in logarithmic frequency.

In this work, we develop a continuous Wilsonian renormalized-flow theory whose central object is an effective nonlinear coupling running across frequency shells and organizing the cascade into a finite transfer branch. The theory is thus formulated directly at the level of the renormalized dynamics that constructs the inertial interval as a finite branch with determinate extent and ultraviolet termination. Its construction follows the Wilsonian logic of successive coarse graining across shells \citep{WilsonKogut1974}, draws on the general logic of diagrammatic closure \citep{Wyld1961} and renormalization in turbulence theory \citep{ForsterNelsonStephen1977,YakhotOrszag1986}, and retains the physical ingredients required in the present setting: dispersion, viscosity, forcing, and resonant interaction topology.

Within this formulation, the inertial interval appears as a plateau of the running flow and acquires the status of an emergent finite branch of spectral transfer. The renormalized flow is non-autonomous in the precise sense that its evolution depends explicitly on the logarithmic distance from the injection scale, and thus encodes the cumulative action of forcing and degradation along the cascade. The renormalized dynamics first determines the existence and extent of the inertial branch, while Kolmogorov--Zakharov spectra arise only as asymptotic constant-flux scaling states realized on that branch. The formulation applies to both capillary and gravity wave turbulence, corresponding respectively to three-wave and four-wave interaction topologies. Monochromatically driven discrete cascades furnish the ultraviolet realization that closes the branch and fixes the topology-dependent exponent structure entering the renormalized flow. In this way, weak wave turbulence is recast as a finite renormalized transfer process in which the inertial branch itself becomes the primary theoretical object and the spectral scaling states emerge as asymptotic structures of that branch. This perspective opens a natural route toward broader nonequilibrium extensions, from nonlocal transfer channels and mixed discrete--continuous cascade dynamics to, at a further level, strongly nonlinear regimes.

\section{Frequency-space formulation}
\label{sec:freq_formulation}

\subsection{Running coupling and resonant-transfer structure}
\label{subsec:running_coupling}

We formulate the theory for weakly nonlinear surface-wave turbulence in finite systems of Newtonian liquids under forcing localized around a prescribed injection scale \(\omega_0\), in the regime realized by laboratory gravity--capillary wave turbulence \citep{FalconMordant2022,Falcon2010,Cobelli2011}. The relevant setting is therefore that of finite forced cascades developing from controlled temporal input and described here, at the level of spectral transfer, in frequency space \citep{ZakharovLvovFalkovich1992,NewellRumpf2011,Nazarenko2011}.

At leading order, the linear wave dynamics is characterized by a dispersion relation of the form
\begin{equation}
\omega(k)\sim k^\beta,
\label{eq:dispersion_scaling}
\end{equation}
with capillary and gravity waves providing the representative limits \(\beta=3/2\) and \(\beta=1/2\), respectively \citep{ZakharovLvovFalkovich1992,Nazarenko2011}. The injection scale \(\omega_0\) then defines the logarithmic frequency coordinate
\begin{equation}
\ell \equiv \ln\left(\frac{\omega}{\omega_0}\right),
\label{eq:ell_def}
\end{equation}
which serves as the natural coarse-graining coordinate of the theory.

The renormalized-flow formulation is organized around a single scale-dependent coupling,
\begin{equation}
g_N(\ell)\equiv g_N\!\left(\ln\frac{\omega}{\omega_0};\,N\right),
\label{eq:gN_def}
\end{equation}
which serves as the central object of the theory, measuring the effective strength of resonant spectral transfer after coarse graining over logarithmic frequency shells and thereby organizing the cascade at the level of its renormalized dynamics. Here \(N\) labels the minimal resonant interaction topology sustaining the cascade. The evolution of \(g_N(\ell)\) defines the renormalized flow, while transfer rates and spectral observables appear as derived quantities. The formal identification of this coupling from the field variables and shell-resolved observables is given in Supplementary Material, Section~S1.

Resonant exchange between modes is constrained, at leading order, by
\begin{equation}
\mathbf{Q}\equiv\sum_{i=1}^{N}\sigma_i\,\mathbf{k}_i=0,
\qquad
\Omega \equiv \sum_{i=1}^{N}\sigma_i\,\omega(\mathbf{k}_i),
\label{eq:resonance_conditions}
\end{equation}
with \(\sigma_i=\pm1\). Momentum conservation ($\mathbf{Q}=0$) is exact, as it follows from translational invariance. By contrast, frequency matching is generally broadened: in finite, weakly nonlinear systems the dispersive relation $\omega(\mathbf{k})$ prevents exact cancellation, so that interactions occur within a quasi-resonant band $|\Omega|\lesssim \Gamma_{nl}$. The minimal interaction order is selected by dispersion: capillary waves admit exact triadic resonances ($N=3$), whereas deep-water gravity waves do not support exact three-wave resonances and transfer energy through effective tetradic interactions ($N=4$) \citep{ZakharovLvovFalkovich1992,Nazarenko2011}. These interaction topologies define distinct universality classes of spectral transfer.

In finite systems and at finite nonlinearity, the resonance manifold is broadened by the finite lifetime of interacting modes. At the scaling level, this broadening may be expressed as
\begin{equation}
|\Omega|
\lesssim
\tau_{\mathrm{nl}}^{-1}(\omega),
\label{eq:broadened_resonance}
\end{equation}
so that near-resonant interactions contribute whenever their detuning remains below the nonlinear linewidth. The exact resonance conditions \eqref{eq:resonance_conditions} thus define the underlying topological class, while the finite interaction time determines the dynamically accessible neighbourhood of that manifold. The corresponding quasi-resonant construction is formalized in Supplementary Material, Section~S2.

In the weakly nonlinear regime, the dominant transfer remains local in scale, in the sense that the interacting frequencies satisfy \(\omega_i\sim\omega\). This locality justifies a shell-wise description in \(\ell\). At the same scaling level, the nonlinear transfer rate takes the form
\begin{equation}
\tau_{\mathrm{nl}}^{-1}(\omega)\sim \omega\,g_N(\ell)^{\,N-2},
\label{eq:tnl_g}
\end{equation}
which expresses the topology-controlled dependence of the transfer time on the running coupling. The locality argument underlying the subsequent RG closure is developed in Supplementary Material, Section~S2.

\subsection{Experimental realization and dimensionless observables}
\label{subsec:exp_realization}

To connect the formal coupling \eqref{eq:gN_def} with measurable quantities, we introduce its realization in the spectral variables used in laboratory surface-wave turbulence. In such systems, spectral transfer is accessed through time-resolved measurements of the free-surface elevation \(\eta(t)\), from which the one-sided elevation power spectral density
\begin{equation}
S_\eta(\omega)\equiv \langle |\hat{\eta}(\omega)|^2\rangle
\label{eq:elevation_psd}
\end{equation}
is obtained by Fourier analysis \citep{Falcon2010,Cobelli2011}. This is the spectral observable used throughout the present formulation.

Denoting by \(a_\omega\) the characteristic elevation amplitude associated with frequency \(\omega\), one has, at the scaling level,
\begin{equation}
a_\omega^2 \sim S_\eta(\omega),
\label{eq:aw_from_psd}
\end{equation}
so that the steepness \(\epsilon(\omega)\sim k(\omega)a_\omega\) satisfies
\begin{equation}
\epsilon(\omega)^2 \sim k(\omega)^2\,S_\eta(\omega).
\label{eq:steepness_psd}
\end{equation}
Accordingly, in the present class of surface-wave experiments, the coupling admits the operational representation
\begin{equation}
g_N^{(\mathrm{exp})}(\ell)\equiv \epsilon(\omega)^2
\sim k(\omega)^2\,S_\eta(\omega),
\label{eq:g_exp}
\end{equation}
which maps the theoretical coupling onto the measured elevation spectrum while preserving its role as the primary RG variable. The formal basis of this identification is given in Supplementary Material, Section~S1.

The localization of forcing around \(\omega_0\) supplies the natural reference scale for a dimensionless description of the cascade. Defining
\begin{equation}
\tilde{\omega}\equiv \frac{\omega}{\omega_0},
\label{eq:omega_tilde}
\end{equation}
one simply has \(\ell=\ln\tilde{\omega}\) by \eqref{eq:ell_def}.

The ultraviolet extent of the cascade is characterized by the reduced cutoff coordinate
\begin{equation}
\bar{\Omega}_\nu\equiv \frac{\omega_K}{\nu k_0^2},
\label{eq:Omega_bar}
\end{equation}
where \(\omega_K\) is the viscous cutoff frequency and \(k_0\) is the wavenumber selected by the forcing frequency through the dispersion relation. The integrated inertial response is measured by
\begin{equation}
\Sigma_{\eta}\equiv \int_{\omega_0}^{\omega_K}S_\eta(\omega)\,d\omega,
\label{eq:Sigma_psd}
\end{equation}
whose reduced form is
\begin{equation}
\bar{\Sigma}_\eta\equiv \frac{\Sigma_{\eta}}{\Lambda_0^2},
\qquad
\Lambda_0\equiv \frac{2\pi}{k_0}.
\label{eq:Sigma_bar}
\end{equation}
Since \(S_\eta(\omega)\) is an elevation spectrum, \(\Sigma_\eta\) has dimensions of length squared, so \(\bar{\Sigma}_\eta\) provides a dimensionless measure of the integrated free-surface response over the inertial interval.

The pair \((\bar{\Omega}_\nu,\bar{\Sigma}_\eta)\) thus furnishes a dimensionless characterization of the cascade, separating its ultraviolet extent from its integrated intensity. In the next section, these observables are combined with \eqref{eq:gN_def} and \eqref{eq:tnl_g} to construct the renormalized flow. The corresponding shell elimination, local beta-function structure, and derivation of the flow equation are given in Supplementary Material, Sections~S3--S5.

\section{Wilsonian coarse-graining: renormalized flow}
\label{sec:RGflow}

\subsection{Frequency-shell coarse graining and beta function}
\label{subsec:beta_function}

Given the logarithmic coordinate \eqref{eq:ell_def} and the running coupling \eqref{eq:gN_def}, the cascade may be recast as a shell-wise coarse-graining problem in frequency space. A Wilsonian step consists in eliminating fluctuations within a narrow shell \([\ell,\ell+d\ell]\) and absorbing their effect into a renormalized interaction acting on the remaining modes. Under the scale-local transfer hypothesis introduced in \S\ref{subsec:running_coupling}, this procedure yields a closed scale evolution for the effective coupling at the same shell, in direct analogy with renormalization-group constructions for turbulent systems \citep{Wyld1961,ForsterNelsonStephen1977,YakhotOrszag1986}. The corresponding shell-elimination argument, the existence of a local beta function, and the origin of its non-autonomous character are developed explicitly in Supplementary Material, Section~S3.

We therefore define the beta function
\begin{equation}
\beta_N(g,\ell)\equiv \frac{dg_N}{d\ell},
\label{eq:beta_def}
\end{equation}
which governs the coarse-grained evolution of the coupling. In the present setting, the flow is necessarily non-autonomous: because the cascade is finite and externally driven, the renormalized dynamics depends not only on the value of \(g_N\) but also explicitly on the running scale \(\ell\). In the language of the Supplementary Material, \(\beta_N(g,\ell)\) is the minimal local beta function compatible with shell-wise elimination in logarithmic frequency; see Section~S3, and in particular Eq.~(S3.6).

\subsection{Structure of the flow}
\label{subsec:flow_structure}

The form of the beta function is constrained by the ingredients already identified in \S\ref{sec:freq_formulation}: scale covariance, forcing localized at the injection scale, cumulative degradation with increasing frequency, and topology-controlled nonlinear transfer. In the Supplementary Material, the local beta-function structure is established in Section~S3, the topology-controlled nonlinear saturation channel is derived diagrammatically in Section~S4, and these ingredients are assembled into a minimal local closure in Section~S5. The resulting flow therefore separates naturally into linear and nonlinear sectors.

First, in the absence of forcing and dissipation, scale covariance implies a linear contribution
\begin{equation}
\left.\frac{dg_N}{d\ell}\right|_{\mathrm{inv}} = c_N g_N,
\label{eq:flow_inv}
\end{equation}
where \(c_N\) is the effective scaling dimension associated with the interaction channel.

Second, the forcing amplitude enters through the injection-scale control parameter \(\mathrm{Re}\). Since forcing acts multiplicatively on the transfer strength, its contribution takes the form
\begin{equation}
\left.\frac{dg_N}{d\ell}\right|_{\mathrm{force}} = \eta_N (\ln \mathrm{Re})\, g_N,
\label{eq:flow_force}
\end{equation}
where \(\eta_N\) measures the topology-dependent shift of the effective linear scaling sector. In the discrete monochromatic realization discussed below and derived in Supplementary Material, Section~S7, one has \(\eta_N=N-2\).

Third, dispersive dephasing and viscous losses progressively reduce transfer efficiency as the cascade advances away from \(\omega_0\). At scaling level, this accumulated degradation is represented by
\begin{equation}
\left.\frac{dg_N}{d\ell}\right|_{\mathrm{diss}} = -\alpha_N \ell\, g_N,
\label{eq:flow_diss}
\end{equation}
with \(\alpha_N>0\). At this stage \(\alpha_N\) remains unspecified: the continuous renormalized flow requires only a monotone local degradation term. Its explicit value is fixed once the ultraviolet realization of the cascade is specified. In the monochromatically driven discrete realization used here, Supplementary Material, Section~S7 yields \(\alpha_N=N/4\).

Finally, nonlinear transfer saturates through the same interaction topology that controls the transfer time in \eqref{eq:tnl_g}. The minimal closure compatible with an \(N\)-wave process is therefore
\begin{equation}
\left.\frac{dg_N}{d\ell}\right|_{\mathrm{nl}} = - B_N g_N^{N-1},
\label{eq:flow_nl}
\end{equation}
with \(B_N>0\). The exponent \(N-1\) is fixed by the leading self-renormalization channel of the effective \(N\)-wave vertex: quadratic for triadic capillary transfer and cubic for effective tetradic gravity transfer. This topology-controlled saturation law is derived diagrammatically in Supplementary Material, Section~S4 and incorporated into the full flow construction in Section~S5.

\subsection{Renormalized flow equation}
\label{subsec:RG_equation}

Combining \eqref{eq:flow_inv}--\eqref{eq:flow_nl} yields the renormalized flow
\begin{equation}
\frac{dg_N}{d\ell}
=
\bigl[c_N+\eta_N \ln \mathrm{Re}-\alpha_N \ell\bigr]\,g_N
-
B_N g_N^{N-1}.
\label{eq:RGflow}
\end{equation}

Equation \eqref{eq:RGflow} is the central dynamical equation of the theory. It defines a non-autonomous Wilsonian flow in logarithmic frequency space, with a linear sector controlled by scale covariance, forcing, and cumulative degradation, and a nonlinear sector fixed by interaction topology. Its formal derivation as a minimal local RG closure is given in Supplementary Material, Section~S5, culminating in Theorem~S5.12.

\subsection{Interpretation}
\label{subsec:RG_interpretation}

The flow \eqref{eq:RGflow} organizes the cascade as a finite renormalized transfer process in logarithmic frequency space. In this description, the running coupling \(g_N(\ell)\) provides a continuous coarse-grained variable that resolves how resonant transfer evolves across shells under the combined action of scale covariance, forcing, cumulative degradation, and topology-controlled nonlinear saturation.

The central consequence is that the inertial regime is determined by the running flow itself and acquires the status of a dynamically selected branch of spectral transfer. Its existence, extent, and termination are fixed by the balance between linear growth, cumulative degradation, and nonlinear saturation. The renormalized-flow description therefore shifts the theoretical focus from spectral scaling taken in isolation to the finite transfer structure that makes such scaling dynamically admissible.

A further distinction is then essential. The continuous flow determines the existence and structure of the inertial branch, whereas its ultraviolet realization remains to be specified separately. In the monochromatically driven case considered below, this ultraviolet sector is realized as a discrete harmonic cascade, which fixes the topology-dependent exponent pair \((\eta_N,\alpha_N)\). This separation between the continuous renormalized flow and its discrete ultraviolet realization is what allows the theory to retain a Wilsonian structure while remaining directly connected to experimentally accessible finite cascades.

In this way, the renormalized dynamics first constructs the finite branch on which asymptotic transport states may be realized. The branch is therefore primary, while its asymptotic scaling states are secondary structures selected within it.
\section{Emergent inertial interval: Kolmogorov--Zakharov spectra}
\label{sec:KZ_plateau}

\subsection{Plateau regime as the inertial interval}
\label{subsec:plateau_regime}

In the present framework, the inertial interval is identified with a quasi-stationary regime of the renormalized flow. Since the flow \eqref{eq:RGflow} is non-autonomous, stationarity is not understood as an exact fixed point, but as a finite range of scales over which the running coupling varies only weakly,
\begin{equation}
\frac{dg_N}{d\ell} \simeq 0.
\label{eq:plateau_condition}
\end{equation}
This condition defines a plateau of the renormalized flow; see also Supplementary Material, Definition~S6.1.

Using \eqref{eq:RGflow}, the plateau condition yields the local balance
\begin{equation}
\bigl[c_N+\eta_N\ln \mathrm{Re}-\alpha_N \ell\bigr]\,g_N
=
B_N\,g_N^{N-1},
\label{eq:plateau_balance}
\end{equation}
or equivalently,
\begin{equation}
g_N^{\,N-2}(\ell)
=
\frac{c_N+\eta_N\ln \mathrm{Re}-\alpha_N \ell}{B_N}.
\label{eq:plateau_branch}
\end{equation}
This is the quasi-stationary branch of the flow; compare Supplementary Material, Proposition~S6.2. Equation \eqref{eq:plateau_branch} shows that the inertial interval is selected dynamically as the finite range of scales over which linear growth and nonlinear saturation remain asymptotically balanced. In physical terms, it is the regime in which the cascade sustains approximately stationary transfer while cumulative degradation has not yet destroyed scale-invariant transport.

\subsection{Asymptotic scaling states}
\label{subsec:asymptotic_scaling}

Within the plateau regime, the nonlinear transfer time \eqref{eq:tnl_g} becomes asymptotically slaved to the slowly varying branch \eqref{eq:plateau_branch}. The renormalized flow therefore establishes the finite branch on which an asymptotic constant-flux transfer state can be realized. Using the experimental realization of the coupling, Eq.~\eqref{eq:g_exp}, together with the standard constant-flux closure of weak wave turbulence, one obtains the asymptotic elevation spectra realized on that branch. In the present formulation, the renormalized dynamics determines the existence and extent of the inertial interval, while the corresponding scaling exponents are selected within that interval by the constant-flux asymptotics. The explicit transformation from the wavenumber-space Kolmogorov--Zakharov spectra to frequency space is given in Supplementary Material, Section~S6.

For capillary waves, corresponding to \(N=3\) and \(\omega\sim k^{3/2}\),
\begin{equation}
S_\eta(\omega)\sim \omega^{-17/6},
\label{eq:KZ_capillary}
\end{equation}
whereas for gravity waves, corresponding to \(N=4\) and \(\omega\sim k^{1/2}\),
\begin{equation}
S_\eta(\omega)\sim \omega^{-4}.
\label{eq:KZ_gravity}
\end{equation}

These exponents coincide with the classical Kolmogorov--Zakharov spectra, but their status is different here. They do not define the inertial interval from the outset. Rather, they appear as asymptotic constant-flux scaling states realized on a finite renormalized branch whose existence and extent are fixed by the flow. The plateau and the KZ spectra therefore play distinct and complementary roles: the former constructs the inertial interval dynamically, whereas the latter specifies the asymptotic scaling structure admissible within it.

\section{Ultraviolet exit and integrated cascade response}
\label{sec:uv_exit}

\subsection{End of the plateau branch}
\label{subsec:plateau_exit}

The inertial interval terminates when the plateau branch \eqref{eq:plateau_branch} ceases to admit a positive real solution for the running coupling. From \eqref{eq:plateau_branch}, this occurs when the numerator vanishes, namely at the scale \(\ell_K\) defined by
\begin{equation}
c_N+\eta_N\ln \mathrm{Re}-\alpha_N \ell_K = 0.
\label{eq:ellK_condition}
\end{equation}
Beyond this point, the quasi-stationary branch ceases to exist and the cascade exits the scale-invariant transfer regime. The corresponding ultraviolet coordinate is therefore
\begin{equation}
\ell_K=
\frac{c_N+\eta_N\ln \mathrm{Re}}{\alpha_N}.
\label{eq:ellK_solution}
\end{equation}
This is the terminal scale of the inertial branch; compare Supplementary Material, Definition~S7.1 and Proposition~S7.2.

Since \(\ell=\ln(\omega/\omega_0)\), the associated cutoff frequency is
\begin{equation}
\omega_K\sim \omega_0\,\exp(\ell_K).
\label{eq:omegaK_general}
\end{equation}
The ultraviolet exit thus appears internally as the terminal point of the renormalized flow and defines the frequency-space analogue of the Kolmogorov cutoff for finite cascades.

\subsection{Scaling of the ultraviolet exit}
\label{subsec:uv_scaling}

Equation \eqref{eq:omegaK_general} may be written as
\begin{equation}
\omega_K
\sim
\omega_0
\exp\!\left(
\frac{c_N}{\alpha_N}
\right)
\mathrm{Re}^{\,\eta_N/\alpha_N}.
\label{eq:omegaK_scaling}
\end{equation}
Up to the non-universal prefactor \(\omega_0\exp(c_N/\alpha_N)\), the ultraviolet cutoff is therefore governed by the exponent ratio \(\eta_N/\alpha_N\).

The status of these two exponents is detailed in Supplementary Material, Section~S7. At the level of the continuous renormalized flow, \(\eta_N\) and \(\alpha_N\) enter as topology- and realization-dependent coefficients; their explicit values are fixed only once the ultraviolet realization of the cascade is specified. Under monochromatic forcing, the cascade is organized as a discrete harmonic ladder, \(\omega_n=n\omega_0\), which provides the physical ultraviolet closure of the inertial branch in finite systems; the corresponding discrete construction is developed in Supplementary Material, Section~S7. The shellwise nonlinear transfer rate then inherits the topology-controlled scaling \(\tau_{nl,n}^{-1}\sim \omega_n g_n^{\,N-2}\), so that the forcing exponent is fixed directly by the \(N\)-wave interaction topology:
\begin{equation}
\eta_N=N-2.
\label{eq:etaN_main}
\end{equation}

A second ingredient is required to determine the degradation exponent. At scaling level, the discrete constant-flux condition constrains the shellwise attenuation of the harmonic ladder but does not by itself uniquely fix the decay law of the shell amplitudes. In the monochromatically driven realization adopted here, we therefore use the minimal topology-consistent amplitude closure \(a_n\sim n^{-N/4}\); this discrete attenuation law and its role in fixing the degradation sector are developed in Supplementary Material, Section~S7. Passing to logarithmic scale, \(\ell=\ln n\), gives \(a_n\sim e^{-(N/4)\ell}\), so that the cumulative degradation coefficient is identified as
\begin{equation}
\alpha_N=\frac{N}{4}.
\label{eq:alphaN_main}
\end{equation}

Accordingly,
\begin{equation}
\omega_K\sim \mathrm{Re}^{\,\eta_N/\alpha_N}.
\label{eq:omegaK_topological}
\end{equation}
This yields
\begin{equation}
\omega_K\sim \mathrm{Re}^{4/3}
\qquad \text{for capillary waves } (N=3),
\label{eq:omegaK_capillary}
\end{equation}
and
\begin{equation}
\omega_K\sim \mathrm{Re}^{2}
\qquad \text{for gravity waves } (N=4).
\label{eq:omegaK_gravity}
\end{equation}

The reduced ultraviolet exit coordinate introduced in \S\ref{subsec:exp_realization},
\begin{equation}
\bar{\Omega}_\nu\equiv \frac{\omega_K}{\nu k_0^2},
\label{eq:Omega_bar_repeat}
\end{equation}
therefore provides the natural RG observable associated with the terminal scale of the inertial branch. It measures the ultraviolet reach of the cascade in units of the viscous decay rate at injection and thereby separates the RG-generated exit scale from the bare dissipative operator fixed at the forcing mode. Since the injection-scale normalization does not alter the topology-controlled exponent structure, \(\bar{\Omega}_\nu\) inherits the same Reynolds scaling as \(\omega_K\),
\begin{equation}
\bar{\Omega}_\nu\sim \mathrm{Re}^{\,\eta_N/\alpha_N}.
\label{eq:Omega_bar_scaling}
\end{equation}
In this sense, the continuous renormalized flow determines the existence and location of the ultraviolet exit, while the monochromatically driven discrete realization fixes the exponent pair that controls its scaling.

\subsection{Integrated inertial response}
\label{subsec:integrated_response}

The ultraviolet extent of the inertial branch does not by itself characterize the finite cascade. A second observable is required to measure the integrated spectral response sustained by that branch. We therefore consider the reduced integrated elevation spectrum introduced in \S\ref{subsec:exp_realization},
\begin{equation}
\bar{\Sigma}_\eta
\equiv
\frac{1}{\Lambda_0^2}
\int_{\omega_0}^{\omega_K} S_\eta(\omega)\,d\omega.
\label{eq:Sigma_eta_repeat}
\end{equation}
In contrast with \(\bar{\Omega}_\nu\), this quantity is normalized only by injection-scale geometric data and therefore probes the cascade response without mixing it with the viscous perturbation that controls the ultraviolet exit.

Unlike the cutoff scale, \(\bar{\Sigma}_\eta\) is not controlled by the ultraviolet end of the inertial interval. Because the elevation spectra relevant here satisfy \(p_N>1\), the integral \eqref{eq:Sigma_eta_repeat} is asymptotically dominated by the infrared edge of the branch, that is, by the matching of the inertial spectrum to the injection scale. In the language of the Supplementary Material, \(\bar{\Sigma}_\eta\) is therefore an infrared response observable, complementary to the ultraviolet coordinate \(\bar{\Omega}_\nu\); see Supplementary Material, Section~S8, in particular Proposition~S8.1 and Remark~S8.1.

The scaling of \(\bar{\Sigma}_\eta\) is fixed by the same topology-controlled nonlinear saturation that governs the renormalized flow, but now through infrared matching at the injection edge of the branch rather than through its ultraviolet termination. At the injection edge of the inertial branch, the effective input is balanced by the first self-renormalization channel of the \(N\)-wave process. This yields a matching coupling
\begin{equation}
g_m(Re)\sim Re^{1/(N-1)},
\label{eq:gm_main}
\end{equation}
while the amplitude of the inertial continuation of the elevation spectrum scales quadratically with that matching response. The reduced integrated response therefore obeys
\begin{equation}
\bar{\Sigma}_\eta(Re)\sim Re^{2/(N-1)},
\label{eq:Sigma_eta_scaling}
\end{equation}
as shown in Supplementary Material, Section~S8.

For the two interaction classes of interest, Eq.~\eqref{eq:Sigma_eta_scaling} gives
\begin{equation}
\bar{\Sigma}_\eta\sim Re
\qquad \text{for capillary waves } (N=3),
\label{eq:Sigma_eta_capillary}
\end{equation}
and
\begin{equation}
\bar{\Sigma}_\eta\sim Re^{2/3}
\qquad \text{for gravity waves } (N=4).
\label{eq:Sigma_eta_gravity}
\end{equation}

The pair \((\bar{\Omega}_\nu,\bar{\Sigma}_\eta)\) thus provides a two-observable characterization of finite wave-turbulent cascades in frequency space. The first measures the ultraviolet reach of the inertial branch; the second measures its integrated response strength. Together they resolve two complementary aspects of the renormalized flow: where the branch ends, and how strongly it is fed from the injection-controlled sector.

\section{Discussion and outlook}
\label{sec:discussion}

\subsection*{Renormalized inertial branch and RG observables}

The present theory recasts weak wave turbulence as a finite renormalized transfer process in frequency space. Its central object is not the spectrum alone, but the running coupling and the finite renormalized branch that it generates. Within this framework, the inertial interval is not introduced as a pre-existing scale-separated window, but constructed dynamically as the quasi-stationary plateau branch of a non-autonomous flow. The condition \(dg_N/d\ell\simeq 0\) does not identify an externally prescribed regime, but selects the finite range of scales over which linear growth, cumulative degradation, and topology-controlled nonlinear saturation balance. In this precise sense, the plateau branch is the inertial interval: it is the dynamical object that replaces the assumed scaling window of Kolmogorov--Zakharov theory.

A direct consequence is that Kolmogorov--Zakharov spectra no longer play a foundational role. They arise as asymptotic constant-flux states realized on the plateau branch, once that branch has been selected by the flow. The logical order is therefore reversed: the renormalized dynamics determines the existence and extent of the inertial interval, and only within that interval do the KZ scaling states emerge. The spectra are thus subordinated to the renormalized branch rather than defining it.

Finite wave-turbulent cascades are correspondingly resolved by two complementary RG observables. The reduced ultraviolet coordinate \(\bar{\Omega}_\nu\) measures the reach of the renormalized branch before viscous arrest, whereas the reduced integrated elevation response \(\bar{\Sigma}_\eta\) measures the spectral strength sustained by that branch through infrared matching at injection. These quantities are not secondary diagnostics of a pre-existing inertial range. They are coordinates of the renormalized flow itself: \(\bar{\Omega}_\nu\) resolves where the branch terminates, and \(\bar{\Sigma}_\eta\) resolves how strongly it is fed. Together they expose the internal organization of the cascade by separating terminal scale selection from cumulative response generation. The inertial branch is thereby resolved not only by its scaling properties, but by its position and weight within the flow.

\subsection*{Constitutive embedding, discrete closure, and outlook}

The framework admits a constitutive embedding without altering its RG structure. Material dependence enters through dispersion, viscous damping, and the injection scale, so that the ultraviolet observable acquires explicit constitutive scaling through the Reynolds number \(Re=U_0\Lambda_0/\nu\), the Bond number \(Bo=\rho g \Lambda_0^2/\sigma\), and the Ohnesorge number \(Oh=\nu/\sqrt{\sigma \Lambda_0/\rho}\), while the integrated response retains a residual constitutive dependence through infrared matching amplitudes (see Supplementary Material, Section~S9). The organization of the flow is thus universal, while its realization is material-specific.

The conceptual leverage of the present framework becomes especially clear under monochromatic forcing. In that case, the cascade does not originate as a pre-formed continuum inertial transfer process, but as a discrete harmonic ladder whose topology and shellwise attenuation close the ultraviolet sector of the renormalized branch and fix the exponent structure governing its ultraviolet exit and integrated response. Discreteness is therefore not a peripheral correction to continuum scaling, but part of the physical mechanism that selects the branch itself. The continuous inertial description emerges only after this ultraviolet closure has been established.

The same perspective sharpens the diagnosis of departures from weak-turbulence behaviour. Whenever the plateau branch is shortened, distorted, or destroyed, the cascade can no longer sustain the asymptotic constant-flux states selected within it. Strong turbulence, finite-size effects, enhanced viscous losses, resonance sparsening, or coherent structures are thus naturally interpreted as mechanisms that deform or suppress the renormalized branch, rather than as perturbations of an otherwise given inertial regime. More broadly, the renormalized-flow formulation places wave turbulence within a Wilsonian nonequilibrium framework while remaining anchored to observables defined directly in frequency space. Natural extensions include nonlocal transfer channels, mixed discrete--continuous cascades, closures beyond weak nonlinearity, and data-driven reconstructions of the running coupling from measured spectra. In all such cases, the relevant question is not only whether a scaling law is observed, but whether the cascade has developed the renormalized branch required for that law to emerge.

\section*{Conclusion}

Weak wave turbulence is thus recast as a finite renormalized transfer process in frequency space. The inertial interval emerges as a plateau branch of the flow, its ultraviolet exit defines the cutoff coordinate, and its infrared matching fixes the integrated response. The pair \((\bar{\Omega}_\nu,\bar{\Sigma}_\eta)\) therefore provides a minimal RG description of finite cascades.

\section*{Acknowledgments}
This work was supported by the Agencia Estatal de Investigación (AEI, Spain) under Grants TED2021-132296B-C52 and CPP2024-011880, and by Fundación BBVA--Fundamentos 2025 (to F.M.), and by the Secretaría de Ciencia, Humanidades, Tecnología e Innovación (SECIHTI, Mexico) under Grant CBF-2025-I-4418 (to J.A.S.).

\bibliography{jfm}

\end{document}